\documentclass{article}
\usepackage{mathrsfs, amsmath, amssymb}
\usepackage[dvipdfm]{graphicx, color}
\usepackage{comment}
\usepackage{cite}

\def\title#1{\begin{center}{\Large\bf #1}\end{center}}
\def\author#1{\vskip 5mm \begin{center}{#1}\end{center}}

\makeatletter
	
	\@addtoreset{equation}{section}
\makeatother
\def\refsec#1{Sec. \ref{sec: #1}}

\def\refapp#1{Appendix. \ref{app: #1}}
\def\refeq#1{(\ref{eq: #1})}

\def\Eq#1{\begin{equation} #1 \end{equation}}
\def\Eqr#1{\begin{eqnarray} #1 \end{eqnarray}}
\def\dfrac#1#2{\displaystyle\frac{#1}{#2}}

\def\p#1{\partial_{#1}}
\def\Ord#1{{\rm O}\left(#1\right)}

\begin{document}


\title{
High frequency limit for gravitational perturbations \\
of cosmological models in modified gravity theories 
}

\author{
	Keiki Saito\footnote{Email address: saitok@post.kek.jp}$^{(a)}$ 
	and Akihiro Ishibashi\footnote{Email address: akihiro@phys.kindai.ac.jp}$^{(b)}$
}

\center{
	$^{(a)}$Department of Particles and Nuclear Physics, \\
	The Graduate University for Advanced Studies (SOKENDAI), \\
	1-1 Oho, Tsukuba, Ibaraki 305-0801, Japan\\
	$^{(b)}$Department of Physics, Kinki University, 
	Higashi-Osaka 577-8502, Japan
}

\abstract{
In general relativity, it has been shown that the effective gravitational 
stress-energy tensor for short-wavelength metric perturbations acts just 
like that for a radiation fluid, and thus, in particular, cannot provide 
any effects that mimic dark energy. 
However, it is far from obvious if this property of the effective 
gravitational stress-energy tensor is a specific nature held only
in the Einstein gravity, or holds also in other theories of gravity.  
In particular, when considering modified gravity theories that involve 
higher order derivative terms, one may expect to have some non-negligible 
effects arising from higher order derivatives of short-wavelength 
perturbations. 
In this paper, we argue this is not the case at least in the
cosmological context. We show that when the background, or
coarse-grained metric averaged over several wavelengths has
FLRW symmetry, the effective gravitational stress-energy tensor for
metric perturbations of a cosmological model in a simple class of $f(R)$
gravity theories, as well as that obtained in the corresponding
scalar-tensor theory, takes a similar form to that in general relativity
and is in fact traceless, hence acting again like a radiation fluid. 
}

\section{Introduction}
\label{sec:Intro}
Our observable universe appears to be homogeneous and 
isotropic on large scales, but highly inhomogeneous on small scales. 
It is therefore considerably interesting to consider whether the local 
inhomogeneities can have any effects on the global dynamics of our 
universe, in particular, any effect that corresponds to a positive
cosmological constant or dark energy. A number of authors have explored
this possibility of explaining the present cosmic accelerating expansion 
by some backreaction effects of the local inhomogeneities 
\cite{R04, KMNR05, KMR06, HS05, MB05, IW06, B06, BLA06, R06, MKMR07, 
		KAF06, K07, W07, BBR08, P08, LABKC09, GMV09, KMM10}. 
Such a backreaction effect may be described in terms of an effective 
stress-energy tensor arising from metric as well as matter perturbations. 

In general relativity, a consistent expansion scheme for short-wavelength 
perturbations and the corresponding effective stress-energy tensor 
were largely developed by Isaacson \cite{I68-1, I68-2}, in which the small 
parameter, say $\epsilon$, corresponds to the amplitude and at the same time 
the wavelength of perturbations. 
Isaacson's expansion scheme is called the high frequency limit 
or the short-wavelength approximation. 
In this expansion, the dominant order of the Einstein equation with respect 
to this parameter $\epsilon$ corresponds to the equations of motion 
for linearized gravitational waves in the ordinary perturbation theory, 
and is in fact divergent as ${\epsilon^{-1}}$.  
The next order of the expansion of the Einstein equation provides 
the Einstein equation for the background metric with 
an effective stress-energy tensor, which is essentially given as 
minus the second-order Einstein tensor averaged over a spacetime region of 
several wavelengths of metric perturbations.  
Since taking a derivative of perturbations corresponds, roughly speaking, 
to multiplying the inverse of the smallness parameter 
(or the inverse of the wavelength of perturbations), the effective 
stress-energy tensor consisting of the square of derivatives of 
the first-order metric perturbations can have large effects on 
the background dynamics. Furthermore it can be shown that 
the effective stress-energy tensor thus constructed is gauge-invariant,  
hence has a physical meaning.  
If the effective stress-energy tensor had a term proportional to the
background spacetime metric, then it would correspond to adding a
cosmological constant to the effective Einstein equations for the
background metric, thereby explaining possible origin of dark energy 
from local inhomogeneities. 
It has been shown, however, that this effective gravitational 
stress-energy tensor is traceless and satisfies the weak energy 
condition, i.e. acts like radiation \cite{B89,GW11}, 
and thus cannot provide any effects that imitate dark energy 
in general relativity. 

However, it is far from obvious if this traceless property of the 
effective gravitational stress-energy tensor is a nature specific only
to the Einstein gravity or is rather a generic property that can hold
also in other types of gravity theories. 
The purpose of this paper is to address this question in a simple, 
concrete model in the cosmological context. 
Among many, one of the simplest of modified theories so far proposed 
is the so called $f(R)$ theory, whose action is a generalization 
of the Einstein-Hilbert action to an arbitrary function, $f(R)$, 
of the scalar curvature $R$. Since $f(R)$ gravity contains higher order 
derivative terms, one can anticipate the effective gravitational 
stress-energy tensor to be generally modified in the high frequency limit.

%
Over the past decade, cosmological implications of $f(R)$ gravity theories 
have extensively been studied especially in the quest of finding an
alternative cosmology to $\Lambda$-CDM model. A viable class of 
the $f(R)$ theories is summarized in 
\cite{HS07, S07, T08, AB07, L09, AGPT07, LB07, FT10, T10}. 
Although it is desirable to examine all these cosmologically favored models,  
in the present paper, 
we will restrict our attention to the simplest model $f(R) = R + cR^2$ 
with an eye to applications to analyses of more generic cases, 
which are left for our future study.  
This model itself is not considered as a cosmologically favored modified 
gravity theory for describing the present accelerating universe, but has 
rather been introduced as a prototype of an inflationary universe model 
by Starobinsky~\cite{Starobinsky80}. 
However, this simple model can be viewed as the leading term truncation 
of a more generic class of $f(R)$ theories that take an analytic form 
with respect to $R$ around the vacuum solution $R=0$ and therefore 
provides, as the first step toward this line of research, 
a good starting point of our analysis. 
It would also be interesting to check whether or not a once-excluded model 
can possibly revive as a cosmologically favored model, due to 
the inclusion of the backreaction effects of local inhomogeneities.

%
It is well-known that $f(R)$ gravity is equivalent to a scalar-tensor 
theory, which contains the coupling of the scalar curvature $R$ to 
a scalar field $\phi$ in a certain way~\cite{FT10, F07, C03, CR10}.  
The Brans-Dicke theory \cite{BD61} is one of the simplest examples. 
%
Therefore our analysis can be performed, in principle, either 
(i) by first translating a given  $f(R)$ theory into the corresponding 
scalar-tensor theory and then inspecting the stress-energy tensor 
for the scalar field $\phi$, or (ii) by directly dealing with metric 
perturbations of the $f(R)$ theory. 
One may expect that the former approach is 
much easier than the latter metric approach, as one has to deal with metric 
perturbations of complicated combinations of the curvature tensors 
in the latter case. 
Nevertheless we will take the both approaches. 
In fact, in the metric approach, by directly taking up perturbations 
of the scalar curvature $R$, the Ricci tensor $R_{ab}$ and 
the Riemann tensor ${R^a}_{bcd}$ involved in a given $f(R)$ theory, 
we can learn how to generalize our present analysis of
a specific class of $f(R)$ gravity to analyses of other, different types
of modified gravity theories that cannot even be translated into
a scalar-tensor theory, such as the Gauss-Bonnet gravity.

%
The theory we consider in the present paper contains higher order 
derivative terms. As in the case of general relativity, we can consider 
short-wavelength metric perturbations with a small parameter $\epsilon$ 
and expand the field equations with respect to $\epsilon$. 
In contrast to the Einstein gravity, the dominant part of the field
equations for this theory is of order $O(\epsilon^{-3})$. 
In $O( \epsilon^{-1})$, we have equations of motion for linearized 
gravitational waves. 
In $\Ord 1 $ we obtain equations for the background metric with a 
source term arising from short-wavelength perturbations. 
This source term contains a number of higher order derivatives of metric
perturbations. However, as one cannot have any meaningful notion of 
stress-energy for gravitational waves in a local sense (at least within 
a wavelength), we have to take a suitable spacetime average over 
several wavelengths. 
Also, since we are interested in backreaction effects on the cosmological 
dynamics, we assume that our background metric takes the form of the 
Friedmann-Lema$\hat{\mbox\i}$tre-Robertson-Walker (FLRW) metric. 
We also impose that in the limit to the Einstein gravity, 
i.e., $f(R) \rightarrow R$ (as $c\rightarrow 0$), the field equations 
for our $f(R)$ theory reduce to those for the Einstein gravity 
in the corresponding order of the expansion parameter. 
At this point, a number of terms that involve higher order derivatives 
of metric perturbations vanish by the spacetime averaging procedure 
and the assumption of the background FLRW symmetry. 
Eventually, besides the terms corresponding to the Isaacson's formula 
in the Einstein gravity, only a few terms that contain higher order 
derivatives of metric perturbations can remain in the effective 
stress-energy tensor for short-wavelength perturbations in our modified 
gravity. Furthermore, the resultant effective stress-energy tensor is 
shown to be traceless as in the Einstein gravity case.

%
We briefly comment on the previous work along the similar line. 
Since the effective stress-energy tensor can be used to measure energy flux 
carried out by gravitational radiation from astrophysical sources, 
such as inspiral binary systems, it can be used to test various modified 
gravity theories by the near future gravitational wave detectors 
\cite{LIGO, VIRGO, KAGRA, SKN01, K_etal_11, P_etal_03, S11, DR03, P03}. 
For this purpose, the effective stress-energy 
tensor for gravitational radiation has been derived by Sopuerta and Yunes \cite{SY09} 
by applying Isaacson's scheme to the field equations of dynamical Chern-Simons 
theory. A more general formalism to compute the effective stress-energy 
tensor from the effective action, which can apply to a wide class of modified 
gravity theories, has been proposed by Stein and Yunes~\cite{SY11}. 
As a concrete example, the formula has been applied to dynamical 
Chern-Simons gravity as well as theories with dynamical scalar fields 
coupled to higher-order curvature invariants. It has been shown that 
in these modified theories the stress-energy tensor for gravitational 
radiations reduces, at future null infinity, to that in the Einstein gravity. 
Berry and Gair~\cite{BG11} also have derived the effective stress-energy 
tensor for gravitational waves in the $f(R)$ gravity which is analytic around 
the vacuum $R=0$. 
However, since the main concern in these studies is mainly to test alternative 
theories in the astrophysical context by using gravitational wave detectors, 
the formulas mentioned just above have been formulated for asymptotically 
flat spacetimes (as energy flux of gravitational waves needs to be evaluated 
at future null infinity) and therefore do not appear to apply to cosmological 
models. In contrast, our analysis will proceed by exploiting the cosmological 
setup that our background metric possesses the FLRW symmetry.

%
In the next section, before going into the effective stress-energy tensor 
in modified gravity theories, we will first briefly summarize the high 
frequency limit in general relativity.  
In \refsec{f(R)}, we consider the high frequency limit in $f(R)$ gravity 
theory. Based on the Isaacson's scheme we expand the field equations for 
$f(R)=R+cR^2$ theory and first derive the general expression 
of the effective stress-energy tensor for gravitational perturbations 
in our $f(R)$ gravity. Then, assuming that our background metric has 
the FLRW symmetry and also that the resulting equations reduce to 
the corresponding equations for the Einstein gravity in the limit 
$c\rightarrow 0$, we see that the effective stress-energy tensor 
whose expression is significantly simplified, is in fact traceless 
as in the Einstein gravity case. 
As briefly mentioned above, when a given $f(R)$ gravity is translated into 
the corresponding scalar-tensor theory, the scalar field $\phi$, 
which expresses an extra-degree of freedom in the $f(R)$ theory, 
possess a non-trivial potential term. [Compare with the earlier work by 
Lee \cite{Lee} on a computation of the effective stress-energy tensor in 
a scalar-tensor theory with vanishing potential term.]
%
In Sec.~\ref{sec: ST}, we will make sure that the effective stress-energy 
tensor in Brans-Dicke theory is consistent with that in our $f(R)$ gravity. 
%
\refsec{summary} is devoted to summary and points to future research.

Our signature convention for $g_{ab}$ is $(-,+,+,+)$. We define the 
Riemann tensor by $R_{abc}{}^d\omega_d= 2\nabla_{[a}\nabla_{b]}\omega_c$ 
and the Ricci tensor by $R_{ab}= R_{acb}{}^c$ as in Wald's book~\cite{Wald84}. 

\section{High frequency limit in general relativity}
\label{sec: GR}

In this section we introduce our notation by recapitulating 
Isaacson's expansion scheme for short-wavelength gravitational 
perturbations in general relativity.

Let $g_{ab}$ be the metric with linear perturbation $h_{ab}$; it is 
described by $g_{ab} = g^{(0)}_{ab} + h_{ab}$ with $g^{(0)}_{ab}$ 
being the background metric including the backreaction from perturbations. 
%
The amplitude of $h_{ab}$ is of order $h_{ab} \sim \Ord\epsilon$ with 
$\epsilon$ being the small parameter, which also corresponds to the wavelength 
$\lambda$ of perturbations compared with the background 
characteristic curvature radius, $L$.  
The order of derivatives of $h_{ab}$ are 
\Eq{
	\nabla_{a_1}\nabla_{a_2} \cdot\cdot\cdot \nabla_{a_m} h_{bc} 
	\sim \Ord{\frac{\epsilon}{(\lambda/L)^m}}
	\sim \Ord{\epsilon^{1 - m}},
}
where $\nabla_a$ denotes the covariant derivative with respect to 
$g^{(0)}_{ab}$, so that $\nabla_a g^{(0)}_{bc} = 0$. 
We may bear in mind perturbations of the form  
$h \sim \epsilon\sin(x/\lambda)$ and $\lambda/L \sim \Ord\epsilon$. 
In what follows we normalize $L \sim 1$. 
The inverse metric takes the form 
$g^{ab} = g^{(0)ab} - h^{ab} + {h^a}_c h^{cb} + \cdot\cdot\cdot$,
where $h^{ab} \equiv g^{(0)ac}g^{(0)bd}h_{cd}$.

%

There is the general relationship between the Ricci tensor of $g_{ab}$ 
and that of $g^{(0)}_{ab}$, namely 
\Eq{
	R_{ab} 
	= R_{ab}[g^{(0)}] 
		+2\nabla_{[c}C^c_{b]a} 
		+2C^c_{d[c}C^d_{b]a} \,, 
} 
where 
$C^a_{bc} \equiv \Gamma^a_{bc} - \Gamma^a_{bc}[g^{(0)}] 
				= g^{ad}\left(\nabla_b g_{dc} 
									+\nabla_c g_{bd}
									-\nabla_d g_{bc}\right)/2$.
Since $R_{ab}$ contains terms such as those schematically expressed as 
$g^{-1}\nabla\nabla g$, $\nabla g^{-1}\nabla g$ and 
$g^{-1}g^{-1}\nabla g\nabla g$, we can find 
\Eq{
	R_{ab}^{(n)}[h] \sim \Ord{\epsilon^{n-2}} \,, 
}
where $n$ is the number of $h_{ab}$ included in $R_{ab}$.  
We also find 
\Eq{
  R^{(n)}[h] \sim G_{ab}^{(n)}[h] \sim \Ord{\epsilon^{n - 2}} \,, 
}
where $R \equiv g^{ab}R_{ab}$, $G_{ab} \equiv R_{ab} - g_{ab}R/2$ 
is the Einstein tensor, and $R^{(n)}[h]$ and $G_{ab}^{(n)}[h]$ 
do not contain $R_{ab}[g^{(0)}]$.

%
The Einstein equation is 
\Eq{
   G_{ab}= 
   R_{ab} - \frac{1}{2}g_{ab}R
	= \kappa^2 T_{ab}^{(0)} \,, 
}
where $\kappa^2 = 8\pi G$ and $T^{(0)}_{ab}$ is 
the stress-energy tensor for the background matter fields. 
%
%
In the following, for simplicity, we focus on metric perturbations 
and ignore perturbations of the matter fields in $T^{(0)}_{ab}$. 
We can find the dominant terms, $\Ord{\epsilon^{-1}}$, of 
the Einstein equation as 
\Eq{
	G_{ab}^{(1)}[h^{\rm }]
	= R_{ab}^{(1)}[h^{\rm }] - \frac{1}{2}g^{(0)}_{ab}R^{(1)}[h^{\rm }]
	= 0 \,,  
\nonumber 
}
or simply 
\Eq{
	R_{ab}^{(1)}[h^{\rm }] = 0 \,. 
\label{eq: O(e^-1)_TT}
}
This is equivalent to the equation for linearized gravitational waves 
in the ordinary perturbation theory. 
Next, in the order of $\Ord 1$, we find 
\Eqr{
    G_{ab}[g^{(0)}]
    &=& \kappa^2 T_{ab}^{(0)} +  \kappa^2 T_{ab}^{\rm eff} \,,
}
where the effective gravitational stress-energy tensor, $T_{ab}^{\rm eff}$, 
is given by 
\Eqr{
\kappa^2 T_{ab}^{\rm eff} &\equiv& - \left< G_{ab}^{(2)}[h^{\rm }]\right> 
\nonumber \\
    	&=& 
	-\left<R_{ab}^{(2)}[h^{\rm }] 
        -\frac{1}{2}g^{(0)}_{ab}g^{(0)cd}R_{cd}^{(2)}[h^{\rm }]\right> 
\nonumber\\ 
	&=& 
\left<\frac{1}{4}\nabla_a h^{{\rm TT}cd}\nabla_b h^{\rm TT}_{cd}\right> 
\,.  
\label{eq: Teff_GR}
}
Here and in the following $\left<\cdot\cdot\cdot\right>$ 
denotes taking a spacetime average over several wavelengths of perturbations.
Here the indices are raised and lowered with $g^{(0)ab}$ and $g^{(0)}_{ab}$.  
For the expression of the third-line, 
the transverse-traceless gauge, $\nabla_ch^c{}_a=0 = h^c{}_c$, 
denoted by ${h}^{TT}$, and \refeq{<R^(2)_ab>_TT} have been 
used (see also \refapp{TT}). 
We can check that $T_{ab}^{\rm eff}$ is traceless, i.e., it acts 
like radiation: 
\Eq{
	\kappa^2 {T^{{\rm eff} a}}_a
	= 0 \,, 
}
from \refeq{O(e^-1)_TT} and \refeq{R^(1)_ab_TT}. 
Thus, in particular, it cannot provide any effects that mimic dark energy 
in general relativity. 
For more mathematically rigorous treatments of short-wavelength perturbations 
and the effective stress-energy tensor, see \cite{B89, GW11}.

The effective stress-energy tensor (\ref{eq: Teff_GR}) can be 
shown to be gauge-invariant \cite{B89}. In fact, 
the expression of the right-hand side of (\ref{eq: Teff_GR}) 
is given by manifestly gauge-invariant part of $h^{\rm TT}_{ab}$. 
For this purpose, one can introduce the polarization tensors $\epsilon_{ab}^{(+,\times)}$, as usual, and decompose the metric perturbation accordingly 
$h^{\rm TT}_{ab}=\epsilon_{ab}^{(+)}h^{(+)} + \epsilon_{ab}^{(\times)}h^{(\times)}$. In the cosmological context, one is concerned with the 
Friedmann-Lema$\hat{\mbox\i}$tre-Robertson-Walker (FLRW) metric, 
\Eq{ 
 ds^2 = -dt^2 + a(t)^2 \gamma_{ij}dx^idx^j 
\label{metric:FLRW}
} 
with $d\sigma^2= \gamma_{ij}dx^idx^j $ being the metric of 
3-dimensional constant curvature space. So, it may be more convenient 
to impose the transverse-traceless condition with respect to this FLRW 
time-slicing, i.e., 
$h^{\rm TT}_{00} = 0, h^{\rm TT}_{0i} = 0, 
 \nabla_a {h^{{\rm TT} a}}_i = 0, {h^{{\rm TT} i}}_i = 0$. This condition 
completely fix the gauge freedom and (\ref{eq: Teff_GR}) is written 
by the gauge invariant variable $h^{\rm TT}_{ij}$ as 
$\left<\frac{1}{4}\nabla_a h^{{\rm TT}ij}\nabla_b h^{\rm TT}_{ij}\right> $.

%

\section{High frequency limit in $f(R)$ gravity}
\label{sec: f(R)}

The general action for the $f(R)$ gravity is given by 
\Eq{
	S = \dfrac{1}{2\kappa^2}\int d^4 x \sqrt{-g}f(R)
			+\int d^4 x \mathcal L_{\rm M} \,,
\label{action:fR}
}
where $\mathcal L_{\rm M}$ is the Lagrangian for matter fields, such as 
perfect fluid in the cosmological context. Varying this action with respect 
to the metric, we have the field equations 
\Eq{
    G_{ab}^{f(R)}
    \equiv G_{ab} +\hat FR_{ab}	-\dfrac{1}{2}g_{ab}\hat f 
            -\nabla_a \nabla_b \hat F 
            + g_{ab}g^{cd}\nabla_c \nabla_d \hat F 
      = \kappa^2 T^{(0)}_{ab} \,,
}
where $\hat f \equiv f - R$, $\hat F \equiv d\hat f/dR$, and 
$T^{(0)}_{ab}$ denotes the matter stress-energy tensor. 

The field equations in the $f(R)$ gravity have terms consisting of 
higher order derivatives of $R$, and the order of those derivatives 
are higher than that of $R$: 
\Eq{
	\nabla_{a_1}\nabla_{a_2} \cdot\cdot\cdot \nabla_{a_m} R^{(n)}[h]
	\sim \Ord{\epsilon^{n - 2 - m}} \,.
}
Therefore it is expected that the effect of the short-wavelength 
approximation would be enhanced. 
In order to see whether this is the case, from now on we restrict our 
attention to the following concrete model 
\Eq{
   f(R) = R + cR^2 \,,
\label{action:R2}
}
where $c$ is a constant. This model has been considered for the first time 
in the context of inflationary universe \cite{Starobinsky80}. 
The field equations are 
\Eq{
 G_{ab}^{f(R)}
 \equiv G_{ab} +2c\left(RR_{ab} -\frac{1}{4}g_{ab}R^2 
               -\nabla_a \nabla_b R + g_{ab}g^{cd}\nabla_c \nabla_d R\right) 
    = \kappa^2 T^{(0)}_{ab} \,.
}

As in Isaacson's formula reviewed in the previous section, we expand the above 
equations with respect to the small parameter $\epsilon$. Then, 
the dominant part is of the order $\Ord{\epsilon^{-3}}$, in which 
we have the following equations 
\Eq{
	\nabla_a \nabla_b R^{(1)}[h]
	-g^{(0)}_{ab}\square R^{(1)}[h]
	= 0 \,. 
}
By contracting with $g^{(0)ab}$, we immediately have 
\Eq{
    \nabla_a \nabla_b R^{(1)}[h] = 0 \,.
\label{eq: O(e^-3)_R^2}
}
Next, for the order $\Ord{\epsilon^{-2}}$, we have 
\Eq{
	R^{(1)}[h]R_{ab}^{(1)}[h]
	-\frac{1}{4}g^{(0)}_{ab}\left(R^{(1)}[h]\right)^2 
	-\nabla_a \nabla_b R^{(2)}[h]
	+g^{(0)}_{ab}\square R^{(2)}[h]
	= 0 \,. 
}
By dotting with $g^{(0)ab}$, we have 
\Eqr{
	\square R^{(2)}[h] 
	&=& 0 \,, 
\\
	\nabla_a \nabla_b R^{(2)}[h]
	&=& R^{(1)}[h]R_{ab}^{(1)}[h]
		-\frac{1}{4}g^{(0)}_{ab}\left(R^{(1)}[h]\right)^2\,. 
\label{eq: O(e^-2)_R^2}
}
Note that since we are working in the short-wavelength approximation, 
we find $g^{(0)ab} R_{ab}^{(1)}[h] = R^{(1)}[h] $ in $\Ord{\epsilon^{-1}}$, 
which is different from calculation in ordinary perturbation theory, 
where in general $g^{(0)ab} R_{ab}^{(1)}[h] \neq R^{(1)}[h]$.
For $\Ord{\epsilon^{-1}}$, we have  
\Eqr{
	&& R_{ab}^{(1)}[h]
		-\frac{1}{2}g^{(0)}_{ab}R^{(1)}[h]
		+2c\left\{\left(R[g^{(0)}] + R^{(2)}[h]\right)R_{ab}^{(1)}[h]
						+R^{(1)}[h]\left(R_{ab}[g^{(0)}] + R_{ab}^{(2)}[h]\right)\right\} \nonumber\\
	&& -\frac{c}{2}\left\{2g^{(0)}_{ab}
								\left(R[g^{(0)}]
										+R^{(2)}[h]\right)R^{(1)}[h]
								+h_{ab}\left(R^{(1)}[h]\right)^2 \right\} \nonumber\\
	&& -2c\left(\nabla_a \nabla_b R^{(3)}[h] 
						-R_{cd}[g^{(0)}]\nabla_a \nabla_b h^{cd}\right) \nonumber\\
	&& +2c\left\{g^{(0)}_{ab}\left(\square R^{(3)}[h] 
											-R_{cd}[g^{(0)}]\square h^{cd}\right)
						-g^{(0)}_{ab}h^{cd}
							\nabla_c \nabla_d R^{(2)}[h]\right\}
		= 0 \,.  
\label{eq: O(e^-1)_R^2}
}
Again, by dotting with $g^{(0)ab}$, we have 
\Eq{
	R^{(1)}[h]
	+\dfrac{c}{2}{h^a}_a\left(R^{(1)}[h]\right)^2
	-6c\left(\square R^{(3)}[h]
				-R_{cd}[g^{(0)}]\square h^{cd}\right)
	+8ch^{ab}\nabla_a \nabla_b R^{(2)}[h]
	= 0 \,. 
\label{eq: trace_O(e^-1)_R^2}
}
In order $\Ord 1$, as in Isaacson's formula in general relativity, 
we have the field equations for the background metric 
with the backreaction source term: 
$G_{ab}^{f(R)}[g^{(0)}] = \kappa^2 T^{(0)}_{ab} + \kappa^2 T^{\rm eff}_{ab}$, 
where  
\Eqr{
 \kappa^2 T^{\rm eff}_{ab}
 &\equiv & -\Biggl<R_{ab}^{(2)}[h]
           -\frac{1}{2}\left(g^{(0)}_{ab}R^{(2)}[h] 
           +h_{ab}R^{(1)}[h]\right) 
\nonumber\\
   && \hspace{16pt} +2c\biggl\{R^{(1)}[h]R_{ab}^{(3)}[h]
     +R[g^{(0)}]R_{ab}^{(2)}[h] 
\nonumber\\
   && \hspace{40pt} +R^{(2)}[h]\left(R_{ab}[g^{(0)}] + R_{ab}^{(2)}[h]\right) 
     +\left(R^{(3)}[h] - h^{cd}R_{cd}[g^{(0)}]\right)R_{ab}^{(1)}[h] \biggr\} 
\nonumber\\
   && \hspace{16pt} 
      -\frac{c}{2}\biggl[g^{(0)}_{ab}
                  \Bigl\{
                         2R^{(1)}[h]R^{(3)}[h]
                        -2R^{(1)}[h]h^{cd}R_{cd}[g^{(0)}]
                        +2R[g^{(0)}]R^{(2)}[h]
                        +\left(R^{(2)}[h]\right)^2 
                  \Bigr\} 
\nonumber\\
   && \hspace{36pt} +2h_{ab}\left(R[g^{(0)}]+R^{(2)}[h]
                            \right)R^{(1)}[h]\biggr] 
\nonumber\\
   && \hspace{16pt} 
      +2c\biggl\{\left(-g^{(0)}_{ab}h^{cd}+h_{ab}g^{(0)cd}\right)
                 \left(\nabla_c \nabla_d R^{(3)}[h]
      -R_{ef}[g^{(0)}]\nabla_c \nabla_d h^{ef}\right) 
\nonumber\\
   && \hspace{42pt} 
      + \left(g^{(0)}_{ab}{h^c}_e h^{ed} - h_{ab}h^{cd}\right)
        \nabla_c \nabla_d R^{(2)}[h]\biggr\} 
      \Biggr> \,.
\label{Teff_R^2_arbitrary} 
} 
This is the expression of the stress-energy tensor for short-wavelength metric perturbations on the generic background metric $g_{ab}^{(0)}$ in our $f(R)$ gravity.


From now on, we consider in the cosmological context. We assume that 
our background is spatially homogeneous and isotropic, that is, our 
background metric possesses the FLRW symmetry and therefore takes 
the form of (\ref{metric:FLRW}). 
Then, thanks to this background symmetry we can explicitly solve 
equations of the form $\nabla_a \nabla_b S(t, \vec x ) = 0$, 
such as \refeq{O(e^-3)_R^2} (see \refapp{solution_S}). 
Equation \refeq{O(e^-3)_R^2} (the equations of motion of 
$\Ord{\epsilon^{-3}}$) is solved to yield 
\Eq{
	R^{(1)}[h]
	= const \,.
}
%
Taking the average, we find 
\Eqr{
	R^{(1)}[h]=const. =\left< const. \right> 
	= \left<R^{(1)}[h]\right> 
        =0 \,. 
\label{R1h=0}
 }
%
Then, the equations \refeq{O(e^-2)_R^2} (those of $\Ord{\epsilon^{-2}}$) 
become 
\Eq{
	\nabla_a \nabla_b R^{(2)}[h]
	= 0 \,. 
\label{DaDbR2h=0}
}
Again using the result in \refapp{solution_S}, we find 
\Eq{
	R^{(2)}[h] \equiv S_1 = const \,. 
\label{R2h=S1}
}
%
%
By using (\ref{R1h=0}) and (\ref{DaDbR2h=0}), 
the equation (\ref{eq: trace_O(e^-1)_R^2}) (of $\Ord{\epsilon^{-1}}$) 
immediately yields  
\Eqr{
	\square R^{(3)}[h]
	-R_{ab}[g^{(0)}]\square h^{ab}
	= 0 \,, 
}
and the equations \refeq{O(e^-1)_R^2} (of $\Ord{\epsilon^{-1}}$) reduce to 
\Eqr{
	\left(1 + 2cR[g^{(0)}] + 2cS_1 \right)R_{ab}^{(1)}[h]
	= 2c\left(\nabla_a \nabla_b R^{(3)}[h] 
						-R_{cd}[g^{(0)}]\nabla_a \nabla_b h^{cd}\right)\,. 
\label{eq: O(e^-1)_R^2_FLRW}
}
The effective stress-energy tensor 
(\ref{Teff_R^2_arbitrary}) is then expressed as 
\Eqr{
	\kappa^2 T^{\rm eff}_{ab}
	&=& -\Biggl<\left(1 + 2cR[g^{(0)}]\right)R_{ab}^{(2)}[h]
								-\frac{1}{2}g^{(0)}_{ab}S_1 \nonumber\\
		&& \hspace{18pt} +2c\left\{\left(R^{(3)}[h] - h^{cd}R_{cd}[g^{(0)}]\right)R_{ab}^{(1)}[h]
												+S_1 \left(R_{ab}[g^{(0)}] + R^{(2)}_{ab}\right)\right\} \nonumber\\
		&& \hspace{18pt} -\frac{c}{2}g^{(0)}_{ab}\left(2R[g^{(0)}] + S_1 \right)S_1 \nonumber\\
		&& \hspace{18pt} +2c\left(-g^{(0)}_{ab}h^{cd}\right)
									\left(\nabla_c \nabla_d R^{(3)}[h]
											-R_{ef}[g^{(0)}]\nabla_c \nabla_d h^{ef}\right)\Biggr> 
\,.
\label{eq: Teff_R^2_proto}
}
Since the effective stress-energy tensor in the $R^2$ model should reduce 
to that in general relativity when $c = 0$, we choose 
$S_1$ ($= R^{(2)}[h]$) to be $0$. 
Then, \refeq{Teff_R^2_proto} becomes 
\Eqr{
	\kappa^2 T^{\rm eff}_{ab}
	&=& -\Biggl<\left(1 + 2cR[g^{(0)}]\right)R_{ab}^{(2)}[h]
						+2c\left(R^{(3)}[h] - h^{cd}R_{cd}[g^{(0)}]\right)R_{ab}^{(1)}[h] \nonumber\\
	&& \hspace{18pt} 
						-g^{(0)}_{ab}\left(1 + 2cR[g^{(0)}]\right)h^{cd}R^{(1)}_{cd}[h]\Biggr> \nonumber\\
	&=& -\Biggl<\left(1 + 2cR[g^{(0)}]\right)R_{ab}^{(2)}[h]
						+2c\left(R^{(3)}[h] - h^{cd}R_{cd}[g^{(0)}]\right)R_{ab}^{(1)}[h]\Biggr>
\,,  
\label{eq: Teff_R^2}
}
where we have used \refeq{O(e^-1)_R^2_FLRW} in the first equality above, 
and 
\Eq{
	\left<h^{cd}R^{(1)}_{cd}[h]\right> = 0
\label{eq: <h^abR^(1)_ab>}
}
in the second equality so as to make the above expression 
compatible with that of general relativity in the $c = 0$ case. 
The expression, (\ref{eq: Teff_R^2}), is our main result of this section. 
%
From $R^{(2)}[h] = 0$ and \refeq{<h^abR^(1)_ab>}, we see 
\Eqr{
	\left<g^{(0)ab}R^{(2)}_{ab}[h]\right>
	&=& \left<R^{(2)}[h] 
					+h^{ab}R^{(1)}_{ab}[h]\right> \nonumber\\
	&=& 0 \,. 
}
Then using this and $R^{(1)}[h] = 0$, we can find that 
$\kappa^2 T^{\rm eff}_{ab}$ is in fact traceless:  
\Eq{
	\kappa^2 {T^{{\rm eff}a}}_a
	= 0 \,.
}

\section{The high frequency limit in scalar-tensor theory}
\label{sec: ST} 

In the previous section, the scalar curvature $R$ 
and the Ricci tensor $R_{ab}$ are taken up directly in the metric formalism of 
the $f(R)$ gravity. 
It is well-known that any $f(R)$ gravity theory is included in
Brans-Dicke theory, which is one of the simplest examples of
scalar-tensor theory \cite{FT10, F07, C03, CR10}. 
In this section, we will see that the results obtained in the previous 
section are indeed consistent with those obtained within the 
corresponding scalar-tensor theory. 

%
The action of Brans-Dicke theory \cite{BD61} is 
\Eq{
 S = \frac{1}{\kappa^2} 
     \int d^4 x \sqrt{-g}\left\{\frac{1}{2}\phi R
                                -\frac{\omega_{\rm BD}}{2\phi} 
                                 \nabla^a \phi\nabla_a \phi
                                -V(\phi)\right\} 
                                +\int d^4 x \mathcal L_{\rm M} \,,
}
where $\omega_{\rm BD}$ is a constant called the Brans-Dicke parameter and 
$\phi$ is a dimensionless scalar field, and 
${\mathcal L}_{\rm M}$ denotes the Lagrangian for matter fields, which can 
in general couple to the metric $g_{ab}$ as well as the scalar field $\phi$.  
Then the equations of motion for $\phi$ and $g_{ab}$ are, respectively, 
obtained as 
\Eqr{
    \square\phi
    +\frac{\phi}{2\omega_{\rm BD}} 
     \left(
           -\frac{\omega_{\rm BD}}{\phi^2}\nabla^a \phi\nabla_a \phi 
           +R -2\p\phi V(\phi) 
     \right)
     &=& \kappa^2 T^{(0)}_{\phi \ ab} \,, 
\\ 
 \phi G_{ab} 
      -\frac{\omega_{\rm BD}}{\phi}
	     \left(\nabla_a \phi\nabla_b \phi
                 - \frac{g_{ab}}{2}\nabla^c \phi\nabla_c \phi 
             \right)
                 -\nabla_a \nabla_b \phi
	+g_{ab} \left( 
                      g^{cd}\nabla_c \nabla_d \phi
					+V(\phi)
                \right)
	&=& \kappa^2 T^{(0)}_{ab} \,,   
}
where $T^{(0)}_{\phi \ ab}$ and $T^{(0)}_{ab}$ are the stress-energy tensor 
for matter fields obtained by taking variations of $\phi$ and $g_{ab}$, 
respectively. 

%
The $f(R)$ gravity of the metric formalism, (\ref{action:fR}), 
can be cast into the form of the above Brans-Dicke theory by setting 
\Eq{
	\phi = F(R) \equiv \frac{df(R)}{dR}\,, 
	\ \ \ \omega_{\rm BD} = 0 \,,  
	\ \ \ V = \frac{F(R)R - f(R)}{2} \,.
}
In this case, as one can find $R - 2\p\phi V= 0$,  
the equations of motion for $\phi$ and $g_{ab}$ just given above become 
respectively 
\Eqr{
	\square\phi 
	-\frac{1}{2\phi}\nabla^a \phi\nabla_a \phi
	&=& \kappa^2 T^{(0)}_{\phi \ ab} \,, 
\\
	G^{\rm ST}_{ab}
	\equiv \phi\left(R_{ab} - \dfrac{1}{2}g_{ab}R\right)
		-\nabla_a \nabla_b \phi
		+g_{ab}\left(g^{cd}\nabla_c \nabla_d \phi + V(\phi)\right)
	&=& \kappa^2 T^{(0)}_{ab} \,.  
}

From now we consider short-wavelength perturbations for $\phi$: 
$\phi = \phi_0 + \delta\phi$. We also assume that there is no coupling 
of matter fields with the second-order derivatives of $\phi$, so that 
there are no non-vanishing terms of order $O(\epsilon^{-1})$ in the 
stress-energy tensor for matter fields. 
Then, the equation of motion for $\phi$ of $\Ord{\epsilon^{-1}}$ is 
\Eq{
	\square\delta\phi = 0 \,,  
\label{eq: square_delta_phi}
}
and the equations of motion for $g_{ab}$ of $\Ord{\epsilon^{-1}}$ are 
\Eq{
	\phi_0 \left(R^{(1)}_{ab}[h] 
        -\frac{1}{2}g^{(0)}_{ab}R^{(1)}[h]\right)
	= \nabla_a \nabla_b \delta\phi 
		-g^{(0)}_{ab}\square\delta\phi \,. 
} 
Contracting with $g^{(0)ab}$, we have 
\Eq{
	R^{(1)}[h]
	= \frac{3}{\phi_0}\square\delta\phi
	= 0 \,,
}
where we have used \refeq{square_delta_phi}. 
From this equation, we can immediately find 
\Eq{
	R^{(1)}_{ab}[h]
	= \frac{1}{\phi_0}\nabla_a \nabla_b \delta\phi \,.
\label{eq: O(e^-1)_ST}
}
The equations of motion in $\Ord 1$ are given by 
$G^{\rm ST}_{ab}[g^{(0)}, \phi_0] = \kappa^2 T^{(0)}_{ab} 
                                   + \kappa^2 T^{\rm eff}_{ab}$, where 
\Eqr{
     \kappa^2 T^{\rm eff}_{ab}
      \equiv -\left<\phi_0 R^{(2)}_{ab}[h] 
               +\delta\phi R^{(1)}_{ab}[h] 
               -g^{(0)}_{ab}h^{cd}\nabla_c \nabla_d \delta\phi 
                \right> \,. 
\label{eq:Teff_ST:general}
}
Here we would like to emphasise that so far we have made 
no assumptions concerning the form of $f(R)$ or the symmetry  
of our background metric $g^{(0)}_{ab}$; 
the above expression, (\ref{eq:Teff_ST:general}), applies to the 
generic $f(R)$ theory with an arbitrary background metric. 

If we restrict the form of $f(R)$ to be (\ref{action:R2}), then 
by inspecting the expansions 
$\phi = \phi_0 + \delta \phi + \cdots$ and 
$F(R) = 1 + 2cR 
		= \left(1 + 2cR[g^{(0)}]\right) 
			+2c\left(R^{(3)}[h] - h^{cd}R_{cd}[g^{(0)}]\right)
			+\cdots$, 
we find 
\Eq{
 \phi_0 = 1+2 c R[g^{(0)}] \,, \quad 
 \delta \phi = 2 c \left(R^{(3)}[h]-h^{cd}R_{cd}[g^{(0)}] \right) \,. 
}
Using these and \refeq{O(e^-1)_ST}, we have 
\Eqr{
     \kappa^2 T^{\rm eff}_{ab}
     = -\left<\left(1 + 2cR[g^{(0)}]\right)R^{(2)}_{ab}[h] 
		  +2c\left( 
                            R^{(3)}[h] -h^{cd}R_{cd}[g^{(0)}]
                     \right)R_{ab}^{(1)}[h]
                  -g^{(0)}_{ab}\phi_0 h^{cd}R^{(1)}_{cd}[h]
             \right> \,. 
\label{eq:Teff_ST:general:metric:fR2}
}

Now we work on the cosmological situation so that 
the background metric has the FLRW symmetry. 
Provided that the limit $c \rightarrow 0$ should reproduce results in 
the case of the Einstein gravity, we finally obtain 
\Eqr{ 
    \kappa^2 T^{\rm eff}_{ab}
	= -\left<\left(1 + 2cR[g^{(0)}]\right)R^{(2)}_{ab}[h] 
	    +2c\left( 
                     R^{(3)}[h] -h^{cd}R_{cd}[g^{(0)}]
               \right)R_{ab}^{(1)}[h] 
             \right> \,,  
\label{eq: Teff_ST}
}
where we have used \refeq{<h^abR^(1)_ab>}, 
derived under the FLRW symmetry in Sec.~\ref{sec: f(R)}.  
We see that the expression \refeq{Teff_ST} above is precisely the same 
as \refeq{Teff_R^2} derived within the metric formalism of the $f(R)$ gravity. 
This verifies our methods of Sec.~\ref{sec: f(R)} for dealing with 
short-wavelength perturbations of the $f(R)$ gravity within the metric 
formalism. 

\section{Summary} 
\label{sec: summary}
We have addressed the effective gravitational stress-energy tensor 
for short-wavelength perturbations in the simple class of $f(R)$ gravity 
of $R^2$ type in the cosmological context. 
As in the Isaacson's formula for the Einstein gravity reviewed 
in Sec.~\ref{sec: GR}, we have obtained the field equations for 
the background metric with a backreaction source term $T^{\rm eff}_{ab}$ 
in order $O(1)$ of the small parameter $\epsilon$. 
Reflecting the fact that our $f(R)$ theory contains higher order derivative 
terms, the source term or the effective stress-energy tensor 
$T^{\rm eff}_{ab}$ takes, as given in (\ref{Teff_R^2_arbitrary}), 
quit a complex form that contains, in principle, terms of fourth order 
derivatives, schematically expressed as 
$\left<\nabla h\nabla h\nabla h\nabla h\right>$. 
The resultant expression, (\ref{Teff_R^2_arbitrary}), of the effective 
stress-energy tensor, in fact, applies to any background 
metric $g^{(0)}_{ab}$; 
Until this point, no symmetry assumption on the background metric has 
been used.  
Then, by imposing that our background has the FLRW symmetry, we have derived 
our effective stress-energy tensor for short-wavelength metric perturbations 
in cosmological models. At this point, thanks to the background FLRW symmetry 
and the spacetime averaging over several wavelengths, the expression of our 
effective stress-energy tensor has been significantly reduced 
to have the simple form, (\ref{eq: Teff_R^2}). 
%
%
We have also shown that the obtained effective stress-energy tensor  
is traceless, so that it acts like a radiation fluid as in the Einstein 
gravity case and thus, in particular, cannot mimic dark energy.

Since any $f(R)$ gravity theory is known to be equivalent to a 
scalar-tensor theory, we have cast our $f(R)$ theory into the corresponding 
scalar-tensor theory. Then, within the scalar-tensor theory, we have
derived the effective stress-energy tensor for short-wavelength
perturbations of the scalar field and checked consistency with the
stress-energy tensor obtained within the metric perturbations of the
original $f(R)$ theory.

Although we have focused on the $R^2$ model especially about the FLRW 
background, in \refsec{ST} we have pushed forward our calculations with
a general $f(R)$ gravity about an arbitrary background
as far as possible, and have not used the property of the $R^2$ model about the
FLRW background, up to (\ref{eq:Teff_ST:general}).  
%
We can immediately note that (\ref{eq:Teff_ST:general}) does not 
involve any terms of fourth order derivatives but has only terms of the  
square of first order derivatives of perturbations $h_{ab}$ and
$\delta\phi$. This result obtained within the framework of the
scalar-tensor theory indicates that the higher order derivatives could 
vanish also in the metric framework of general $f(R)$ gravity theory
for a generic background. However, to see whether this is indeed the case 
needs further involved calculation, and is beyond the scope of this 
paper. This is left open for future study.

%
Our formulas derived in \refsec{f(R)} deal directly with the scalar curvature 
$R$ and the Ricci tensor $R_{ab}$, and therefore should be able to apply 
to similar analyses of other modified gravity theories which contain 
higher order curvature terms composed of $R$, $R_{ab}$, and ${R^a}_{bcd}$ 
and which cannot even be cast in the form of a scalar-tensor theory. 
It would be interesting to consider an extension of our present work 
to a wide class of modified gravity theories with high-rank curvatures. 
%

%
%

%
%
%
%
%

\section*{Acknowledgements}

We would like to thank Hideo Kodama, Hirotaka Yoshino, Robert M. Wald, 
and Stephen R. Green for useful discussions and comments.
This work was supported in part by a Grant-in-Aid for JSPS Fellows 
under Grant No. 23-4465 (KS) and by the JSPS Grant-in-Aid 
for Scientific Research (C)No. 22540299 (AI) and by the Barcelona 
Supercomputing Center (BSC) under Grant No. AECT-2012-2-0005.

\appendix
\section{Perturbation formulas in Transverse-Traceless gauge}
\label{app: TT}

The conditions of transverse-traceless waves are 
$\nabla_a {h^{{\rm TT} a}}_b = {h^{{\rm TT} a}}_a = 0$. 
The Ricci tensors for $h^{\rm TT}_{ab}$ are
\Eqr{
C^{a(1)}_{ab}[h^{\rm TT}] &=& 0 \,, 
\\
R^{(1)}_{ab}[h^{\rm TT}] &=& \nabla_c C^{c(1)}_{ba}[h^{\rm TT}] \,,
\nonumber\\									&=& -\frac{1}{2}\square h^{\rm TT}_{ab} \label{eq: R^(1)_ab_TT}\\
	R^{(2)}_{ab}[h^{\rm TT}] &=& 2\nabla_{[c}C^{c(2)}_{b]a}[h^{\rm TT}] 
											-C^{c(1)}_{db}[h^{\rm TT}]C^{d(1)}_{ca}[h^{\rm TT}] 
\nonumber\\
 &=& -\frac{1}{2}h^{{\rm TT} cd}									\nabla_c \Bigl(\nabla_b h^{\rm TT}_{da}
																+\nabla_a h^{\rm TT}_{bd}
																-\nabla_d h^{\rm TT}_{ba}\Bigr) 
\nonumber\\										&& +\frac{1}{2}
												\nabla_b \left\{h^{{\rm TT} cd}
																		\Bigl(\nabla_c h^{\rm TT}_{da}
																				+\nabla_a h^{\rm TT}_{cd}
																				-\nabla_d h^{\rm TT}_{ca}\Bigr)\right\}
											\nonumber\\
										&& -\frac{1}{4}\Bigl(\nabla_d {h^{{\rm TT} c}}_b
																	+\nabla_b {{h^{\rm TT}}_d}^c
																	-\nabla^c h^{\rm TT}_{db}\Bigr)
															\Bigl(\nabla_c {h^{{\rm TT} d}}_a
																	+\nabla_a {{h^{\rm TT}}_c}^d
																	-\nabla^d h^{\rm TT}_{ca}\Bigr) \,, 
}
and the scalar curvatures for $h^{\rm TT}_{ab}$ are 
\Eqr{
	R^{(1)}[h^{\rm TT}] &=& g^{(0)ab}R_{ab}^{(1)}[h^{\rm TT}] \nonumber\\
								&=& 0 \,, \\
	R^{(2)}[h^{\rm TT}] &=& g^{(0)ab}R_{ab}^{(2)}[h^{\rm TT}] 
									 -h^{{\rm TT} ab}R_{ab}^{(1)}[h^{\rm TT}] \nonumber\\
					&=& \frac{3}{4}\nabla^a h^{{\rm TT} bc}\nabla_a h^{\rm TT}_{bc}
							-\frac{1}{2}\nabla^a h^{{\rm TT} bc}\nabla_c h^{\rm TT}_{ab}
							+h^{{\rm TT} ab}\square h^{\rm TT}_{ab} \,, 
}
where we have used 
$[\nabla_a , \nabla_b]h^{\rm TT}_{cd} 
	= R_{c \ ab}^{\ e}[g^{(0)}]h^{\rm TT}_{ed}
		+ R_{d \ ab}^{\ \, e}[g^{(0)}]h^{\rm TT}_{ce}
	= \Ord\epsilon$. 
From \refeq{O(e^-1)_TT} and \refeq{R^(1)_ab_TT}, 
we find $\square h^{\rm TT}_{ab} = 0$. Using these we find  
\Eqr{
	\left<R^{(2)}_{ab}[h^{\rm TT}]\right>
	&=& -\frac{1}{4}\nabla_a h^{{\rm TT} cd}\nabla_b h^{\rm TT}_{cd}, \label{eq: <R^(2)_ab>_TT} \\
	\left<R^{(2)}[h^{\rm TT}]\right>
	&=& 0 \,. 
}

\section{Solution of $\nabla_a \nabla_b S(t, \vec x ) = 0$}
\label{app: solution_S}

We solve the equation 
\Eq{
	\nabla_a \nabla_b S(t, \vec x )
	= 0 \,,
\label{eq: ddS=0}
}
such as \refeq{O(e^-3)_R^2}, in the FLRW background spacetime: 
\Eq{
	ds^2
	= -dt^2
		+ a^2 (t) \gamma_{ij}dx^i dx^j \,. 
}
Since we are interested in the expanding universe, in what follows 
we assume that the scale factor $a(t)$ is dynamical, i.e., 
$a(t) \neq const$.  
The $(0, 0)$, $(0, i)$ and $(i, j)$ components of (\ref{eq: ddS=0}) are 
\Eqr{
	\ddot S(t, \vec x ) &=& 0, \label{eq: 00}\\
	\p i \dot S(t, \vec x ) - \frac{\dot a(t)}{a(t)}\p i S(t, \vec x ) &=& 0, \label{eq: 0i}\\
	\p i \p j S(t, \vec x ) - a(t)\dot a(t)\gamma_{ij}\dot S(t, \vec x ) 
	- \Gamma^{k (3)}_{ij}\p k S &=& 0, \label{eq: ij}
}
where the {\it dot} denotes the derivative with respect to the cosmic time. 
The solution of \refeq{00} is 
\Eq{
	S(t, \vec x ) = c_1 (\vec x ) t + c_2 (\vec x ) \,,
} 
where $c_1 (\vec x )$ and $c_2 (\vec x )$ are arbitrary functions of $\vec x $. 
%
%
%
Since $\dot a \neq 0$, Equation \refeq{0i} becomes 
\Eq{
	\left(\frac{a(t)}{\dot a(t)} - t\right)\p i c_1 (\vec x) 
	= \p i c_2 (\vec x) \,. 
}
Therefore, either 
\Eq{ 
	c_1, c_2 = const.
}
or $\partial_i c_2 /\partial_i c_1= const.$. In the latter case, 
by shifting $t \rightarrow t- const.$, we have 
${\dot a}/a = 1/t$, and therefore have $a(t)\propto t$; the behavior 
of the background FLRW universe is determined. This is, in our present 
context, too restrictive, and for this reason, we should take 
the former case; $c_1, c_2=const. $. 
%
Then, equation \refeq{ij} for $i = j$ becomes 
\Eq{
	a(t)\dot a(t)\gamma_{ij}c_1
	= 0 \,,
}
which immediately implies 
\Eq{
	c_1 = 0 \,. 
}
Therefore we find the solution of \refeq{ddS=0} to be 
\Eq{
	S = const.
}


\end{document}